\documentclass[%
 reprint,
 superscriptaddress,
nofootinbib,
 amsmath,amssymb,
aps,
prl,
]{revtex4-2}

\usepackage[ruled,linesnumbered]{algorithm2e}
\usepackage{amsfonts, amsmath, amssymb, amsthm}
\usepackage{braket}
\usepackage{graphicx}
\usepackage{dcolumn}
\usepackage{bm}
\usepackage{extpfeil}
\usepackage{mathtools}
\usepackage{subcaption}   

\newtheorem{example}[]{Example}

\def\Z2{$\mathbb{Z}_2$}

\begin{document}

\preprint{APS/123-QED}
\title{\Z2 Lattice Gauge Theory on Non-trivial Topology and Its Quantum Simulation}

\author{Jiaqi Hu}
\altaffiliation{These authors contributed equally to this work.} 
\affiliation{Center for Intelligent and Networked Systems, Department of Automation, Tsinghua University, Beijing 100084, China}

\author{Shu Tian}
\altaffiliation{These authors contributed equally to this work.} 
\affiliation{Department of Physics \& State Key Laboratory of Surface Physics, Fudan University, Shanghai 200438, China}

\author{Xiaopeng Cui}%
\email{xpclove@126.com}
\affiliation{Department of Physics \& State Key Laboratory of Surface Physics, Fudan University, Shanghai 200438, China}

\author{Rebing Wu}%
\email{rbwu@tsinghua.edu.cn}
\affiliation{Center for Intelligent and Networked Systems, Department of Automation, Tsinghua University, Beijing 100084, China}

\author{Man-Hong Yung}%
\affiliation{Shenzhen Institute for Quantum Science and Engineering, Southern University of Science and Technology, Shenzhen, 518055, China}
\affiliation{International Quantum Academy, Shenzhen, 518048, China}
\affiliation{Guangdong Provincial Key Laboratory of Quantum Science and Engineering, Southern University of Science and Technology, Shenzhen, 518055, China}
\affiliation{Shenzhen Key Laboratory of Quantum Science and Engineering, Southern University of Science and Technology, Shenzhen, 518055, China}
\date{\today}

\author{Yu Shi}%
\email{yu\_shi@ustc.edu.cn}
\affiliation{Wilczek Quantum Center, Shanghai Institute for Advanced Studies, University of Science and Technology of China, Shanghai 201315, China}
\begin{abstract}
Wegner duality is essential for \Z2 lattice gauge theory, yet the duality on non-trivial topologies has remained implicit.
We extend Wegner duality to arbitrary topology and dimension, obtaining  a new class  of Ising models, in which topology is encoded in non-local domain-wall patterns. 
Without the overhead of gauge constraints, simulating this model on an $L\times L$ torus requires only $L^2$ qubits with two-body couplings, halving the conventional four-body coupled $2L^2$ qubits,  
enabling full experimental realization of \Z2 lattice gauge theory on near-term devices.
\end{abstract}

\maketitle


\nocite{*}

\textit{Introduction.}
Lattice gauge theories (LGTs) originate from Wilson's non-perturbative framework for quantum chromodynamics~\cite{Wilson74} and the subsequent Hamiltonian formulation by Kogut and Susskind~\cite{Kogut75}. Duality of LGTs\cite{Savit80} was introduced by Wegner,  with the map between  (2+1)D $\mathbb{Z}_2$ lattice gauge theory ($\mathbb{Z}_2$ LGT) and the Ising model~\cite{Wegner71}, later generalized to $\mathbb{Z}_N$, $\mathrm{U}(1)$ \cite{Fradkin78} and $\mathrm{SU}(N)$\cite{Mathur_2016} LGTs. The physical significance of $\mathbb{Z}_2$ LGT was further elevated by  topological orders in  a variety of systems such as  fractional quantum Hall effect~\cite{Laughlin83}, string-net condensation~\cite{Wen90A, Wen90B, Wen05, Wen07book}, superconductivity\cite{hansson_superconductors_2004}, and quantum error correction codes, e.g. the toric code ~\cite{Kitaev97Toric}.

This  trend has stimulated considerable recent experimental activity of  quantum simulation of $\mathbb{Z}_2$ LGTs  demonstrated on both classical emulators\cite{Cui2020, Ding2022, yamamoto_quantum_2022, fukushima_violation_2023, Wang2025, Wauters2025} and on near-term quantum processing platforms\cite{Wiese2013, bender_digital_2018, Surace2020, Lumia2022, Fromm2024, Charles2024, Mueller2025}. In a complementary direction, the original Wegner duality has been applied inversely to map challenging optimization problems, encoded as Ising models, onto experimentally more accessible $\mathbb{Z}_2$ LGT Hamiltonians for quantum annealing~\cite{Zoller15} and quantum inspired classical optimization algorithms\cite{Wang2024}.

Wegner's duality indicates a promising approach for quantum simulation of \Z2LGTs\cite{yamamoto_real-time_2021}, i.e. simulating the  dual Ising model instead of directly dealing with  \Z2LGT, which possesses  high-order couplings and gauge constraints. However, the most  discussed \Z2LGT
on a toroidal geometry exhibits fourfold degeneracy in
its ground states, a characteristic absent in its dual Ising
model. This distinction arises because Wegner's duality is exact on simply connected spaces. Its extension to non-trivial topologies has remained implicit for quantum simulation. Although pioneering works\cite{Fradkin78, Savit80, yao_dualization_2024} have discussed \Z2LGT on the torus, there has been a lack of explicit spin model for the Wegner duality of \Z2LGT on non-trivial topologies,  forming a major obstacle for experimental studies. 

In this Letter, we extend the  Hamiltonian form of Wegner duality and its use in  quantum simulation to non-trivial topologies that is compatible for near-term intermediate scale quantum (NISQ) devices. Using a homological framework similar to CSS quantum error correction codes\cite{Calderbank_1996, Kitaev97Toric,Bombin_2007}, we show that local dynamics of the $\mathbb{Z}_2$ LGT is dominated by fluctuations of closed strings, while the ground state degeneracy corresponds to the homology group. In the dual model, spins correspond to  local closed strings in the original model,  and there are nonlocal couplings with the  auxiliary spins,  different values of which  correspond  to different topological sectors.  This gives rise to a \emph{Sectorial Ising (SI) model}, which enables gauge-free quantum simulation of an $L\times L$ torus with $L^2$ qubits per sector, halving the $2 L^2$ qubit requirement in  conventional approaches at the cost of multiple independent runs. Since SI model requires less qubits than the toric code model, it stands as a new resource-efficient model for quantum simulation of topological order.

\textit{Closed String Representation of \Z2LGT.}
Let $\Sigma$, $\Lambda$ and V be the plaquettes, links and vertices of an $L\times L$ torus,
and on each link we place a qubit. The star and plaquette operators detect \Z2 electric charge and magnetic vortices, respectively  
\begin{equation} \label{eq:AvBp}
    A_v = \prod_{\ell\in \delta v} X_\ell,\quad B_\sigma = \prod_{\ell\in\partial \sigma} Z_\ell,
\end{equation}
where $\partial\sigma\subset \Lambda$ are the boundary links of plaquette $\sigma$,
and $\delta v\subset \Lambda$ are the  links   adjacent to vertex $v$.   \Z2LGT is formulated as
\begin{align}
    H = -\sum_{\sigma\in\Sigma} \prod_{\ell\in\partial \sigma} Z_\ell + g \sum_{\ell\in \Lambda} X_\ell,\label{eq:z2ham}
\end{align}
of which   $\{A_v\}_{v\in V}$ is  the gauge symmetry operators. 

States of the \Z2LGT one-to-one map  to superpositions of closed strings, giving rise to a \emph{closed string representation}. 
We choose gauge fixing as the +1 common eigenspace of the $\{A_v\}$ operators. 
The closed strings are excited by Wilson string operators $\prod_{\ell\in w} Z_\ell$ for each loop $w$ on the primal lattice, forming a basis of the gauge-fixed Hilbert space, as proved in Appendix A. The deconfined phase ($g = 0$) ground state of \Z2LGT in the trivial topological sector, $\ket{g_{00}} \sim \sum_{w} \prod_{\ell\in w}Z_\ell \ket{+}$, is the superposition of all contractible closed strings, an example of Wen's string-net condensation\cite{Wen05}. States in non-trivial topological sectors are obtained by $\tilde{X}_\gamma = \prod_{\ell\in\gamma} X_\ell, \gamma = \gamma_1,\gamma_2$.

\textit{Wegner duality on non-trivial topology.}
The conventional Wegner duality, in absence of topological non-triviality, maps the $B_\sigma$
to $\tilde{X}_\sigma$ of a qubit on plaquette $\sigma$, and $X_\ell$ to $\prod_{\sigma\in\delta \ell}\tilde{Z}_\ell$
where $\delta\ell \subset \Lambda$ are the plaquettes adjacent to $\ell$. In case of non-trivial topology, 
we place a qubit not only on each plaquette, but also on each independent non-contractible loop, e.g. $\mathcal{C} = \{\gamma_1, \gamma_2\}$ where $\gamma_1$ and $\gamma_2$ are independent non-contractible loops on the torus. Then we transform to the closed string representation by
\begin{align}
    \tilde{X}_{\sigma} &= \prod_{\ell\in \partial \sigma} Z_\ell, \label{eq:repr-transfo-z}\\
    X_\ell &= \Bigl(\prod_{\sigma\in\delta\ell} \tilde{Z}_\sigma\Bigr) \Bigl(\prod_{\gamma\owns \ell} \tilde{Z}_\gamma\Bigr)\label{eq:repr-transfo-x}.
\end{align}
This formulation extends the Wegner duality with the topological term $\tilde{Z}_\gamma$ corresponding to non-trivial loop $\gamma$ entering the link $\ell$.

\textit{General Topology and the Hodge Perspective.}
\begin{figure}[t]
\centering
\includegraphics[width=\linewidth]{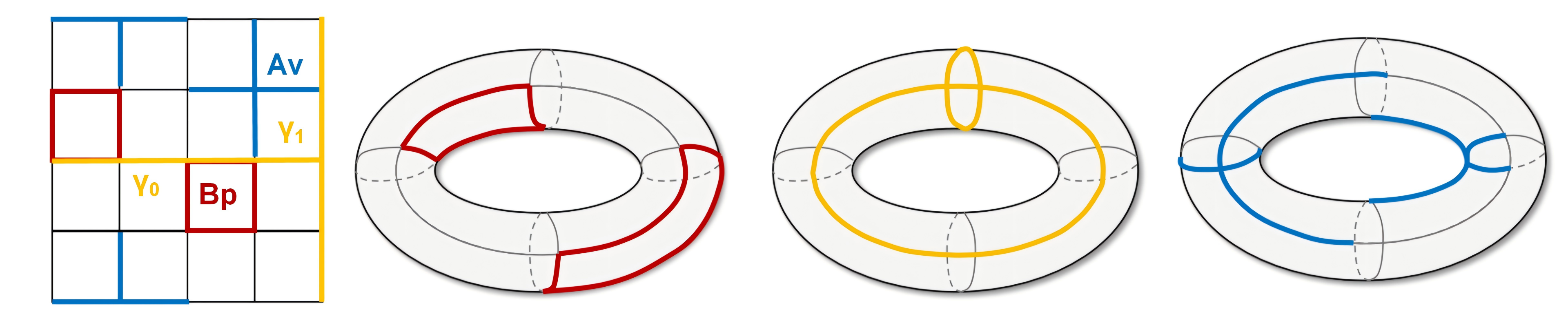}
\caption{Hodge decomposition of the string Hilbert space into contractible closed strings (red), harmonic fields (gold), and gauge transformations (blue),  mathematically described by $\mathrm{im}(\partial)$, $\mathrm{ker}(\Delta)$, and $\mathrm{im}(\delta)$.}
\label{fig:Hodge-decomposition}
\end{figure}
The aforementioned transformation is extended to general topologies in arbitrary dimension. To capture the commuting algebra of the star ($A_v$) and plaquette ($B_p$) operators in \Z2LGT, we employ combinatorial Hodge theory, natural frameworks to decompose a field into source-free, curl-free and harmonic components.

Consider a general cell complex (e.g. a lattice) 
\begin{equation}
    \mathbb{Z}_2^\Sigma \underset{\delta}{\overset{\partial}{\rightleftharpoons}} \mathbb{Z}_2^\Lambda \underset{\delta}{\overset{\partial}{\rightleftharpoons}} \mathbb{Z}_2^V, \partial^2 = 0,\delta^2 = 0.
\end{equation}
where $V$, $\Lambda$, and $\Sigma$ are three finite sets, $\partial$ and $\delta$ being the boundary and coboundary operators with relation $\delta=\partial^T$. Here 
we implicitly assume the isomorphism between chain groups and cochain groups, which allows the boundary operator $\partial$ and the coboundary operator $\delta$ to act in the same spaces $\mathbb{Z}_2^\Sigma$, $\mathbb{Z}_2^\Lambda$, and $\mathbb{Z}_2^V$. For the mathematical construction that explicitly handles this subtlety, please refer to Appendix B. 

On each element $\ell\in\Lambda$ resides a spin, with all possible spin configurations (up/down) described by $\mathbb{Z}_2^\Lambda$.
At each element $v\in V$ there can be a \Z2 electric charge , with all charge configurations described by $\mathbb{Z}_2^V$. The star operators $\{A_v\}$ to detect these charges are described by the coboundary operator $\delta$.
At each element $\sigma\in \Sigma$ there can be a \Z2 magnetic flux, with all flux configurations described by $\mathbb{Z}_2^\Sigma$. The plaquette operators $\{B_p\}$ to detect these fluxes are described by the boundary operator $\partial$.

On non-trivial topologies, star and plaquette operators---encoding divergence and curl of $\mathbb{Z}_2$ LGT fields---cannot uniquely determine the field (unlike Euclidean space). The missing harmonic component is captured by the Hodge-Laplace operator $\Delta = \partial \delta + \delta \partial$, defining $\mathcal{T} = \ker \Delta$ as the subspace that is both curl- and source-free, encoding purely topological properties. The combinatorial Hodge theorem gives an orthogonal decomposition of the Hilbert space:
\begin{equation} \label{eq:hodge_decomp_general}
\mathbb{Z}_2^\Lambda = \operatorname{im}(\delta) \oplus \mathcal{T} \oplus \operatorname{im}(\partial).
\end{equation}
This decomposition separates the physical degrees of freedom: $\operatorname{im}(\delta)$ constitutes the gauge degrees of freedom of the \Z2LGT,
$\mathcal{T}$ comprises harmonic fields with a one-to-one mapping to the topological sectors, responsible for the ground state degeneracy, 
and $\operatorname{im}(\partial)$ encodes the the dynamics of contractible strings, which are local degrees of freedom. 

By projecting the \Z2LGT Hamiltonian to the gauge-invariant subspace $\mathcal{T}\oplus \mathrm{im}(\partial)$ via
the duality transformation \eqref{eq:repr-transfo-z}, \eqref{eq:repr-transfo-x}, we define the gauge-free \emph{sectorial Ising (SI) model}.

\textit{Sectorial Ising Model.}
\begin{figure}[t]
    \centering
    \includegraphics[width=\linewidth]{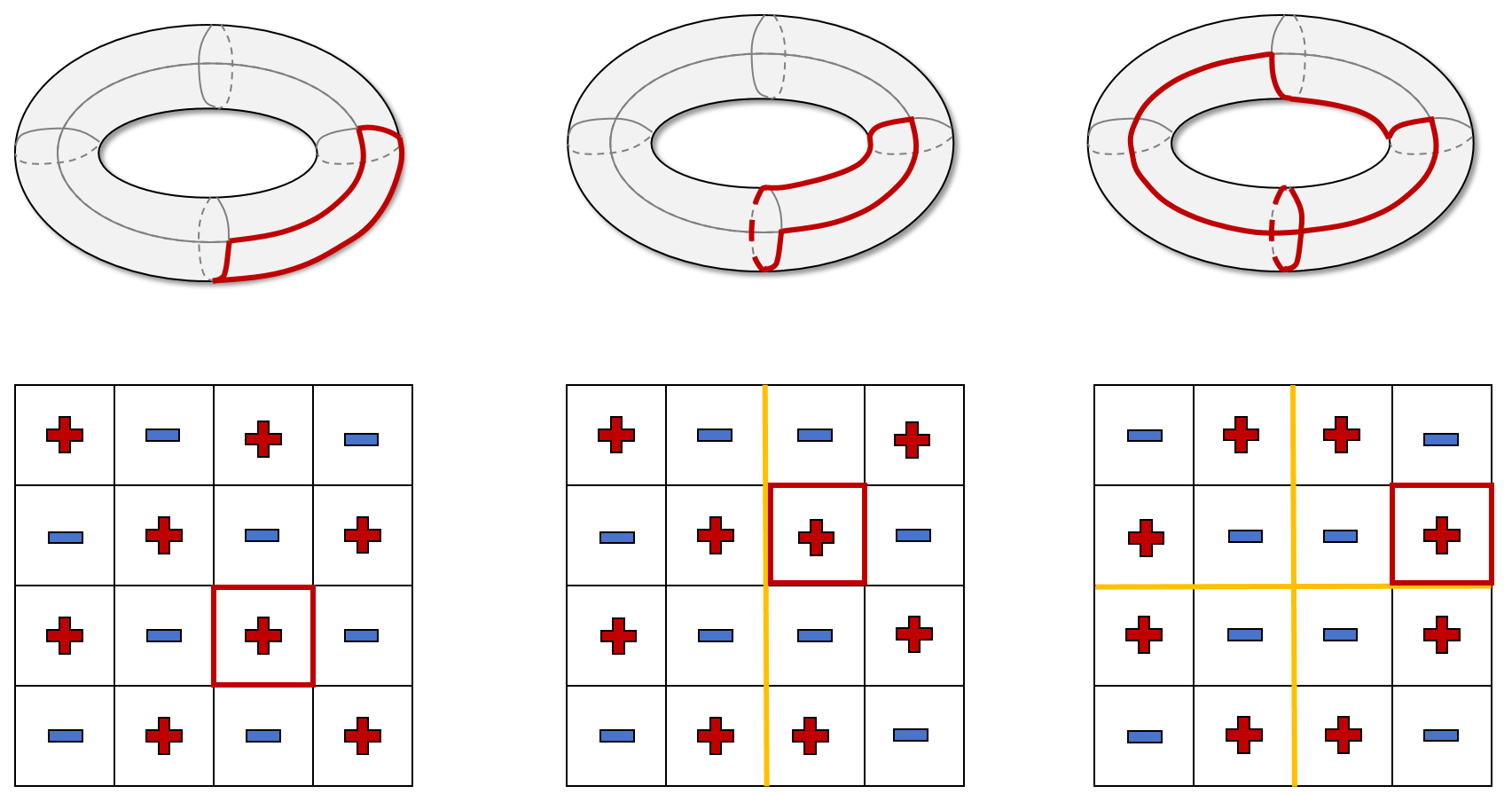}
    \caption{Ground state of SI model at $g=0$: each sector (left/mid/right) has a unique ground state set by domain walls; left column with trivial topology reduces to the Ising model. Dual spin (red border) maps to a closed string in the corresponding \Z2LGT sector.}
    \label{fig:sIM}
\end{figure}
Applying the dual transformation \eqref{eq:repr-transfo-z} and \eqref{eq:repr-transfo-x},
we obtain the Hamiltonian dual to \Z2LGT on non-trivial topology
\begin{equation}\label{eq:SI}
    H' = -\sum_{\sigma\in\Sigma} \tilde{X}_\sigma + 
    g\sum_{\ell\in\Lambda} \prod_{\sigma\in\delta\ell} \tilde{Z}_\sigma \cdot \prod_{\gamma\owns \ell} \tilde{Z}_\gamma.
\end{equation}
This Hamiltonian takes the transverse-field Ising form, but with nonlocal couplings to auxiliary qubits $\{\tilde{Z}_\gamma\}$ that encode the topological sector. We refer to this as the Sectorial Ising (SI) model, where the dual of a non-contractible loop in the \Z2LGT is an auxilliary topological qubit. In the trivial sector ($\tilde{Z}_\gamma = +1$), it reduces to the standard Ising model.

Note that the plaquette operators in the original \Z2LGT are dependent. States with $\prod_{\sigma\in\Sigma} \tilde{X}_\sigma = +1$ symmetry
describes the dynamics of the \Z2LGT. As $[\prod_{\sigma\in \Sigma} \tilde{X}_\sigma, H'] = 0$,
it suffices to ensure the symmetry of the initial state to be $\prod_{\sigma\in \Sigma} \tilde{X}_\sigma = +1$.

The redundancy in the plaquette operators reveals topology characterized by non-contractible surfaces other than non-contractible loops. It can be eliminated by depicting a plaquette $\sigma_0$ and let $\Sigma^* = \Sigma - \{\sigma_0\}$,  establishing 
\begin{align}
    H &= -H_M + g H_E, \label{eq:dual}\\
    H_M &= \sum_{\sigma\in\Sigma^*} \tilde{X}_\sigma + \prod_{\sigma\in\Sigma^*} \tilde{X}_\sigma,\label{eq:dual-M}\\
    H_E &= \sum_{\ell\in\Lambda}\prod_{\substack{\sigma\in\delta\ell\cap\Sigma^*}} \tilde{Z}_\sigma \cdot \prod_{\gamma\in\mathcal{C}_\ell} \tilde{Z}_\gamma.\label{eq:dual-E}
\end{align}
Origin of the global product term in \eqref{eq:dual-M} is mathematically described by the Betti number and high-order cohomology groups, as shown in Appendix B.

\textit{The Sectorial Ising Model as an Emergent Topological Phenomena.}
The SI model has two key features: (1) it describes the same wavefunction as \Z2 gauge field theory
through a topologically induced construction, and (2)
it shares features like degenerate ground states with topological materials,
but crucially differs in its handling of sector transitions.

Remarkably, the operators $\tilde{Z}_\gamma$ commute with  \eqref{eq:SI}.
This inherent symmetry allows for further reduction of $\mathcal{T}$ in \eqref{eq:hodge_decomp_general}, leaving
only $\mathrm{im}(\partial)$ degrees of freedom.
This induces the Hamiltonian
\begin{align}
    H(\{z_\gamma\}) = -\sum_{\sigma\in \Sigma} \tilde{X}_\sigma + g \sum_{\ell\in \Lambda} \prod_{\gamma\owns \ell} (-1)^{{(1-z_\gamma)}/2} \prod_{\sigma\in\delta\ell} \tilde{Z}_\sigma.
\end{align}

\begin{itemize}
    \item \textbf{Trivial sector:}
    When $\tilde{Z}_\gamma = +1$ for all $\gamma \in \mathcal{C}$,
    the Hamiltonian reduces to a transverse-field Ising model.
    This essentially recovers the result given by Wegner, Fradkin, and Susskind.
    
    \item \textbf{Non-trivial sectors:}
    For configurations where $\tilde{Z}_{\gamma_i} = -1$ on a set of non-contractible loops $\{\gamma_i\}$,
    the Hamiltonian describes
    a phase transition from a ferromagnetic state to a paramagnetic state with domain walls,
    which constitutes a generalized Ising model.
    Each distinct configuration of $\{\tilde{Z}_\gamma = \pm 1\}$
    defines a unique topological sector of the $\mathbb{Z}_2$ gauge theory.
\end{itemize}
In the ferromagnetic phase, the dominant energy contribution comes from the magnetic term,
rendering the sign of the electric term physically irrelevant. 
Consequently, SI model exhibits identical behavior
to the Ising model under ferromagnetic ordering.
This demonstrates how topologically protected phenomena 
in the $\mathbb{Z}_2$ gauge theory manifest in its dual Ising model.

The SI model possesses an intrinsic gauge freedom: for a fixed configuration of $\{\tilde{Z}_\gamma = \pm 1\}$, non-contractible loops can be selected in numerous distinct ways. Each choice yields a Hamiltonian with different coefficients, yet all share identical Hilbert spaces---analogous to a group admitting multiple generator choices. This freedom can be experimentally advantageous, allowing selection of higher-quality qubits to form the non-contractible loops and enhance fidelity.

The SI model eliminates gauge degrees of freedom but preserves closed string dynamics of the \Z2LGT, maintaining string-net condensation with topological order and sector classification. While the toric code was historically deemed the simplest such model---typically requiring $2L^2$ qubits on an $L\times L$ lattice---the SI model achieves the same physics with up to three-body couplings at the cost of gauge fixing, fewer than the four-body terms required for the toric code.

Since a transverse Ising model can be mapped to a higher-dimensional classical Ising model, the SI model can similarly correspond to the following model:
\begin{align}
    \label{classical si model}
    H_{csi}(\{z_\gamma\}) = -K\sum_{\sigma\in \Sigma}\left(s_{\sigma,1}s_{\sigma,N} + \sum_{n=1}^{N-1} s_{\sigma, n}s_{\sigma, n+1}\right) \notag \\
    + g\Delta\tau \sum_{n=1}^N \sum_{\ell\in\Lambda}\prod_{\gamma\owns \ell} (-1)^{{(1-z_\gamma)}/2} 
    \prod_{\sigma\in \delta\ell}s_{\sigma, n}.
\end{align}
where imaginary time direction has been divided into $N$ time slices of width $\Delta\tau$ and $\exp(-2K) = \tanh(\Delta\tau) $\cite{Rieger_1994}.
Using techniques such as Monte Carlo methods for simulating Ising models,
we can deal with larger lattice scales on current computer hardware and also provide guidance for future simulations on quantum computers.

\textit{Examples of \Z2LGT with non-trivial topology.}
Topological structures are essential for the performance of topological quantum codes. Similarly, by employing the language of chain complexes, we are able to characterize the \Z2LGT Hamiltonian and its dual on arbitrary topology. As shown in the following examples, however, not all non-trivial topological properties(e.g. non-zero Betti numbers in example 1 and 3) provide protection. 
\begin{example}\label{example:tetrahedron}
For \Z2LGT on a tetrahedron,
\begin{align*}
    H_M &= Z_1 Z_2 Z_4 + Z_2 Z_3 Z_5 + Z_1 Z_3 Z_6 + Z_4 Z_5 Z_6,\label{eq:tetrahedron-z}\\
    H_E &= X_1 + X_2 + X_3 + X_4 + X_5 + X_6,
\end{align*}
The dual SI Hamiltonian is
\begin{align*}
    H_M &= \tilde{X}_1 + \tilde{X}_2 + \tilde{X}_3 + \tilde{X}_1 \tilde{X}_2 \tilde{X}_3,\\
    H_E &= \tilde{Z}_1 \tilde{Z}_2 + \tilde{Z}_2 \tilde{Z}_3 + \tilde{Z}_1 \tilde{Z}_3 + \tilde{Z}_1 + \tilde{Z}_2 + \tilde{Z}_3.
\end{align*}
\end{example}
Note that if $Z_4 Z_5 Z_6$ is removed, the global product term $\tilde{X}_1 \tilde{X}_2 \tilde{X}_3$ also vanishes since the $H_M$ topology becomes the disk, and the dual Hamiltonian also becomes the Ising model. From the mesh perspective, this phenomenon describes a bulk-edge effect. By Stokes theorem
\begin{displaymath}
    \prod_{\ell\in\partial_2 \sigma_0} Z_\ell = \prod_{\sigma\in\Sigma^*} \prod_{\ell\in\partial_2 \sigma} Z_\ell,
\end{displaymath}
the right hand side is $\prod_{\sigma\in\Sigma^*} \tilde{X}_\sigma$. The global product term describes $\mathbb{Z}_2$ magnetic flux living on the boundary of the open surface $\Sigma^*$. The total $\mathbb{Z}_2$ magnetic flux is conserved on the surface and its boundary 

\begin{example}
The dual SI Hamiltonian of \Z2LGT on a $2\times 2$ torus is
\begin{align*}
    H_M &= \tilde{X}_{10} + \tilde{X}_{01} + \tilde{X}_{11} + \tilde{X}_{01} \tilde{X}_{10} \tilde{X}_{11},\\
    H_E &= 
    \tilde{Z}_{\gamma_0} \tilde{Z}_{10} + \tilde{Z}_{\gamma_0} \tilde{Z}_{01} \tilde{Z}_{11} +
    \tilde{Z}_{10} + \tilde{Z}_{01} \tilde{Z}_{11} + \notag\\
    &\qquad\tilde{Z}_{\gamma_1} \tilde{Z}_{01} + \tilde{Z}_{\gamma_1} \tilde{Z}_{10} \tilde{Z}_{11} +
    \tilde{Z}_{01} + \tilde{Z}_{10} \tilde{Z}_{11}.
\end{align*}
The torus has non-trivial fundamental group whose generators $\gamma_0, \gamma_1$ are explicit in the closed string Hamiltonian.
\end{example}
\begin{example} 
Consider two tetrahedrons sharing a common link, as shown in fig \ref{fig:double-tetrahedron}. The primal and dual $\mathbb{Z}_2$LGT Hamiltonians are
\begin{align*}
    H_M &= Z_1 Z_2 Z_6 + Z_1 Z_3 Z_5 + Z_2 Z_3 Z_4 + Z_4 Z_5 Z_6 + \notag\\
    &\qquad Z_6 Z_7 Z_8 + Z_6 Z_{10} Z_{11} + Z_7 Z_9 Z_{10} + Z_8 Z_9 Z_{11}.\\
    H_M' &= \tilde{X}_1 + \tilde{X}_2 + \tilde{X}_3 + \tilde{X}_1 \tilde{X}_2 \tilde{X}_3 + \notag\\
    &\qquad\tilde{X}_4 + \tilde{X}_5 + \tilde{X}_6 + \tilde{X}_4 \tilde{X}_5 \tilde{X}_6.
\end{align*}
\begin{figure}
    \centering
    \includegraphics[width=0.8\linewidth]{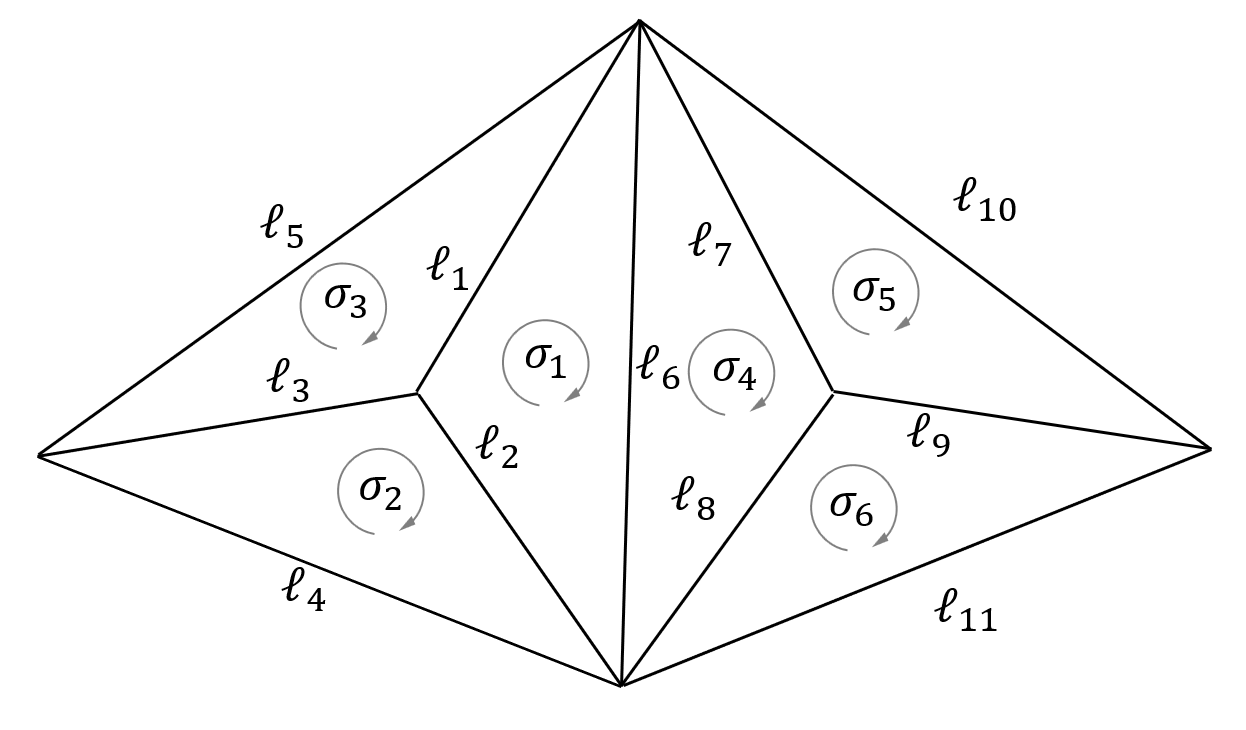}
    \caption{Two tetrahedrons sharing a link}
    \label{fig:double-tetrahedron}
\end{figure}
\end{example}
This example shows topological properties of a Hamiltonian can be different from the geometry that define the system, as the mesh in Fig. \ref{fig:double-tetrahedron} is planner but the Hamiltonian has the topology of two tetrahedrons union.

\textit{Digital Simulation of \Z2LGT on a Torus}
The duality between \Z2LGT and sectorial Ising models provides a gauge-free model for quantum simulation. 
A \Z2LGT model can be simulated with minimal resources by performing independent quantum simulation for each sector of the sectorial Ising model.
Thus \Z2LGT on the $L\times L$ torus can be simulated by $L^2$ qubits per sector, requiring only two-body couplings.

\begin{figure}[t!]
    \centering
    \includegraphics[width=\linewidth]{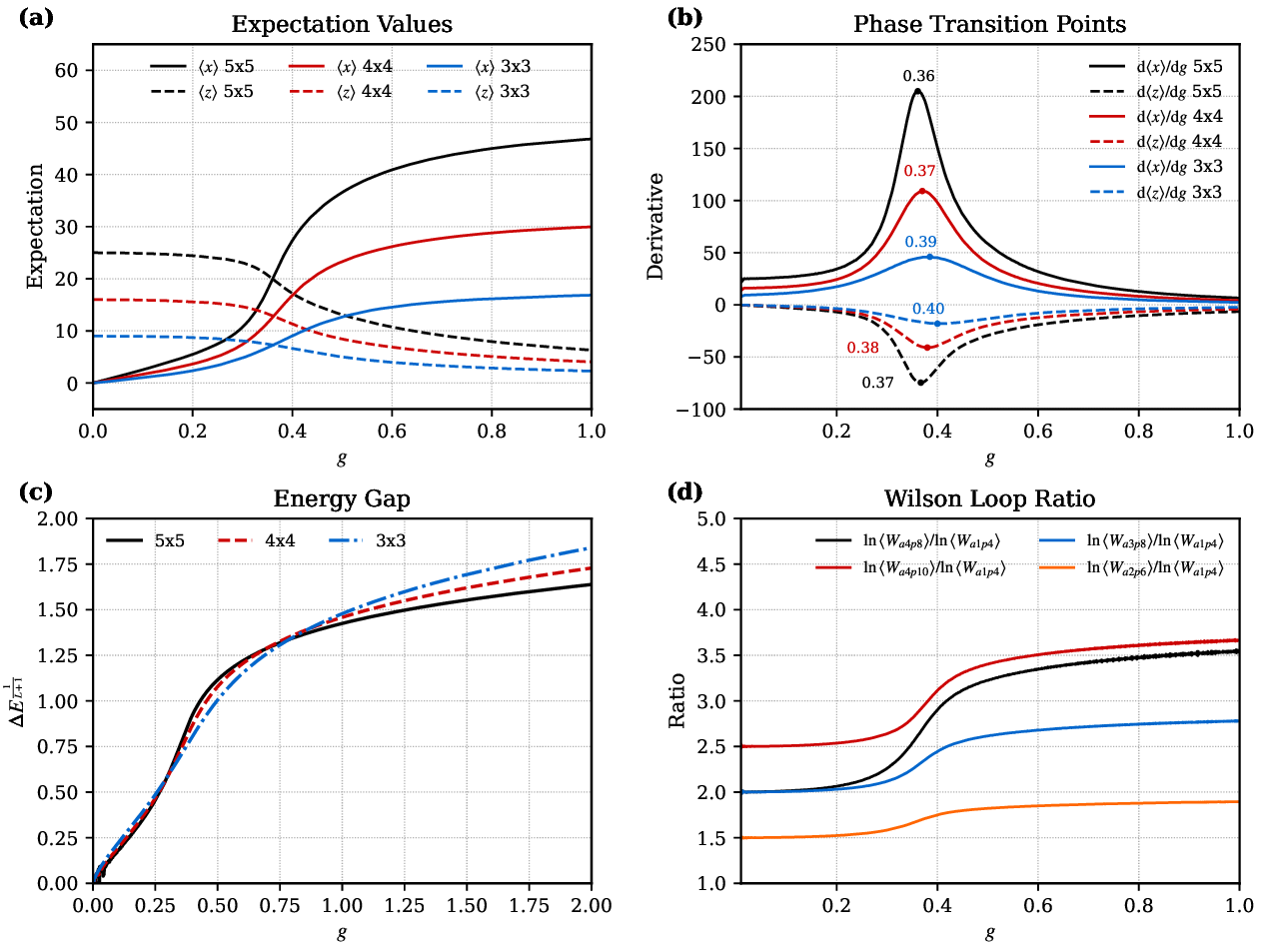}
    \caption{Quantum phase transition simulation of the model \eqref{eq:dual}-\eqref{eq:dual-E} on $L\times L$ tori. (a) Energy expectations. (b) Phase transition points $g_c$. (c) Gap $\Delta E$. (d) Wilson loops $\langle W_C\rangle$.
    100,000 Trotter steps with step-size 0.01 controls $\langle x\rangle = \langle H_E\rangle$ and $\langle z\rangle = \langle H_M\rangle$ absolute errors below $10^{-3}$. Gradient is calculated by central differential method of step 0.01, with moving average filter window 0.01.}
    \label{fig:results}
\end{figure}

Quantum phase transition of \Z2LGT on torus of different sizes are observed, as shown in Fig. \ref{fig:results}.
The critical point $g_c$ is measured as extrema points of $\mathrm{d}\langle x\rangle / \mathrm{d}g$ and $\mathrm{d} \langle z\rangle / \mathrm{d}g$, which are obtained by central difference of step 0.01 and moving average filter of window size 0.01, showing $g_c\approx 0.36$ on $5\times 5$ torus and increases when lattice size decreases. 
The energy gap  $\Delta E$  between ground states $\ket{g_{00}}$ and $\ket{g_{01}}$ agrees with $\Delta E \propto g^{\frac{1}{L+1}}$ near the origin\cite{Sachdev_2018}. 
The expectation of Wilson loops obeys perimeter law and area law in the deconfined and confined phases respectively. The simulation results are consistent with the theoretical prediction of \Z2LGT\cite{Sachdev_2018}.

\textit{Conclusion.}
We extend Wegner duality of the \Z2LGT to non-trivial topology, formulating the sectorial Ising (SI) model together with its classical form, that is highly compatible for near-term quantum simulators. Topological protection emerges naturally from the correspondence between topological sectors of the \Z2LGT and domain-wall configurations of the SI model. This model enables resource-efficient quantum simulation of \Z2LGTs, reducing the simulation cost of \Z2LGT on an $L\times L$ torus from four-body coupled $2L^2$ qubits to the two-body coupled $L^2$ qubits of the sectorial Ising model.  

\textit{Acknowledgments.}
This work is supported by the Innovation Program for Quantum Science and Technology (Grants No. 2021ZD0300200) and the National Natural Science Foundation of China (Grant No. 62173201 and 12075059). 

\textit{Author Contributions.}
Jiaqi Hu and Shu Tian contributed equally to this work. 
Jiaqi Hu carried out the mathematics and Trotter simulation.
Shu Tian conceived the sectorial Ising model and related physical picture.
Xiaopeng Cui provide the insight to reduce gauge degrees of freedom of \Z2LGT simulation on the torus.
Rebing Wu, Manhong Yung and Yu Shi supervised the project. 
All authors discussed the results and contributed to the final manuscript.

\bibliography{ref}

\newpage

\appendix

\section{Appendix A. Completeness of Closed String Representation Proven by Euler's Formula}
We show that all states of the \Z2LGT within the gauge fixed $\{A_v = +1\}_{v\in V}$ subspace $\mathcal{H}_G$ are superposition of closed strings. We show with explicit construction that (i) the closed strings are in $\mathcal{H}_G$, and (ii) $\mathcal{H}_G$ contains only the closed strings due to the constraint of dimensionality.

Note that $\ket{+}\in \mathcal{H}_G$. Moreover, any loop $w$ intersects even number of links adjacent to a vertex,  
$$[\prod_{\ell\in w} Z_\ell, A_v] = 0.$$ 
Thus $\prod_{\ell\in w} Z_\ell \ket{+} \in \mathcal{H}_G$, proving (i).

On a closed surface $S$ with $g_S$ genus, there are $|\Sigma| - 1$ independent plaquette borders due to closed topology of $S$, and there are two independent non-contractible loops across each genus, rendering 
$$\dim(\mathcal{H}_G) \geq 2^{|\Sigma| + 2 g_S - 1}.$$ 
On the other hand, a closed surface has $|\Lambda|$ qubits and  $|V| - 1$ independent gauge operators, demanding 
$$\mathrm{dim}(\mathcal{H}_G) \leq 2^{|\Lambda| - |V| + 1}.$$
By Euler's formula $|\Sigma| - |\Lambda| + |V| = 2 - 2 g_S$,
we show $\mathrm{dim}(\mathcal{H}_G) = 2^{|\Sigma| + 2 g_S - 1}$, proving (ii).

\section{Appendix B. Homological Structure of $2\times 2$ Torus}
\begin{figure}[t]
\includegraphics[width=0.5\linewidth]{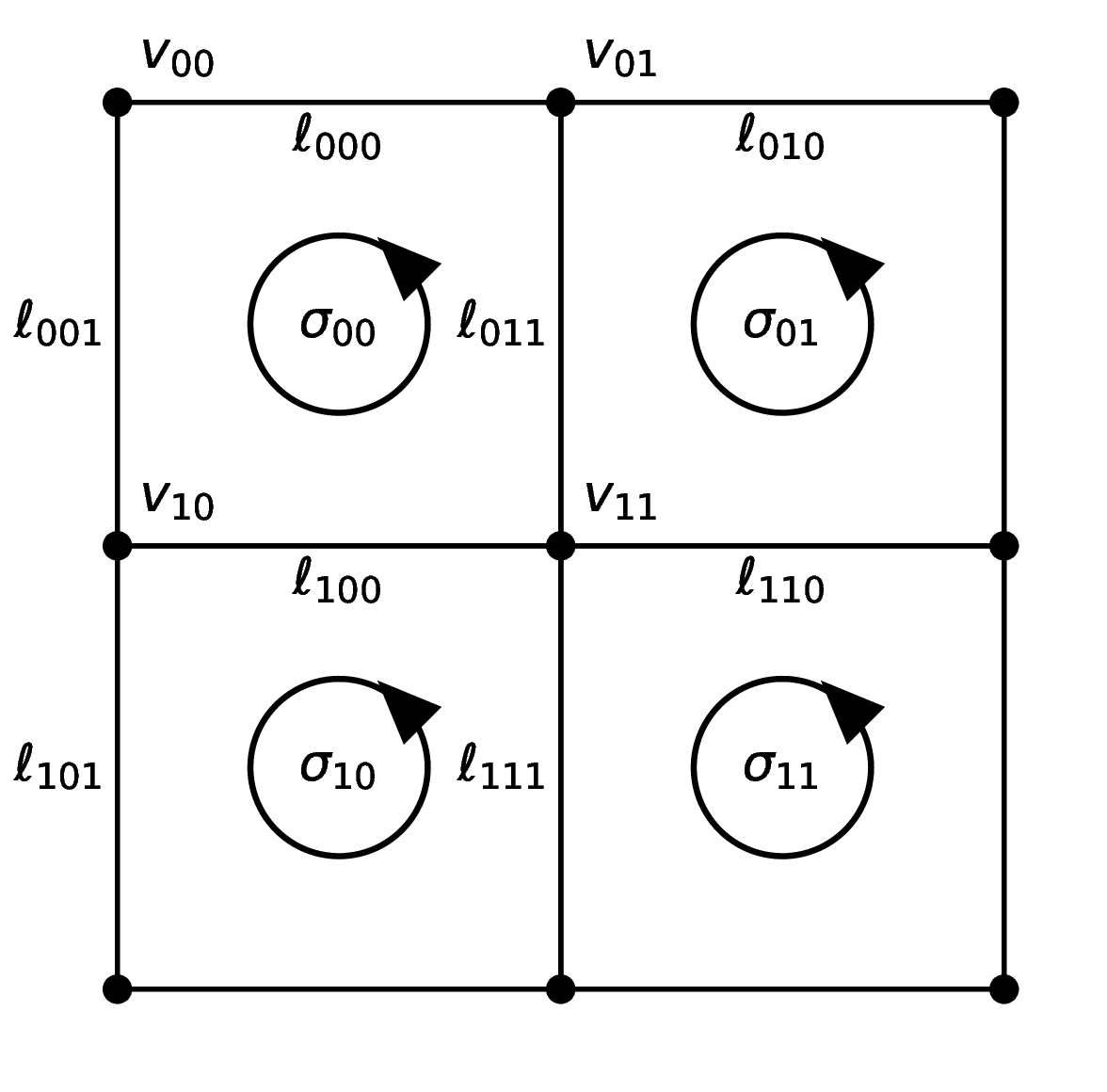}
\caption{A $2\times 2$ lattice with periodic boundary condition.}
\label{fig:2x2-torus}
\end{figure}

In this appendix we investigate the homological structure of the $2\times 2$ torus as shown in Fig. \ref{fig:2x2-torus}.
Let the set of all plaquettes, links and vertices be
$\Sigma = \{\sigma_{00}, \sigma_{01}, \sigma_{10}, \sigma_{11}\}$,
$\Lambda = \{\ell_{000}, \ell_{001}, \ell_{010}, \ell_{011}, \ell_{100}, \ell_{101}, \ell_{110}, \ell_{111}\}$ and
$V = \{v_{00}, v_{01}, v_{10}, v_{11}\}$. We define chain groups as the free vector space over $\mathbb{Z}_2$,
\begin{align*}
C_0 &= \sum_{v\in V} c_v v, \quad C_1 = \sum_{\ell\in \Lambda} c_\ell \ell,\quad C_2 = \sum_{\sigma\in \Sigma} c_\sigma \sigma,
\end{align*}
where $c_v, c_\ell, c_\sigma\in\mathbb{Z}_2$. They are organized by the boundary operators $\partial_2, \partial_1$, or $\partial$ when determined by context, 
\begin{align*}
    \partial_2 \sigma &= \sum_{\ell \text{ is boundary of }\sigma}\ell,\\
    \partial_1 \ell &= \sum_{v \text{ is boundary of } \ell} v.
\end{align*}
e.g. $\partial \sigma_{00} = \ell_{000}+\ell_{001}+\ell_{100}+\ell_{011}$, $\partial \ell_{000} = v_{00}+v_{01}$. 
An important property of the boundary operator is
\begin{displaymath}\label{eq:homology}
    \partial \circ \partial = 0,
\end{displaymath}
e.g.
$\partial_1\circ\partial_2 (\sigma_{00}) = \partial_1 (\ell_{000} +\ell_{001} +\ell_{100} +\ell_{011})\\
= (v_{00} + v_{01})+ (v_{01}+ v_{00}) +(v_{00}+ v_{10}) +(v_{10}+ v_{00})
= 0.$
In other words, the border of some plaquettes always intersects an even number of links adjacent to a vertex. This establishes the chain complex
\begin{displaymath}\label{eq:chain-complex}
    C_2 \stackrel{\partial_2}{\longrightarrow}
    C_1 \stackrel{\partial_1}{\longrightarrow}
    C_0, \quad \partial_1\circ\partial_2 = 0.
\end{displaymath}

The topology is well characterized by the chain complex. The boundary group $B_i$ and the cycle group $Z_i$ are defined as
\begin{align*}
    B_i &= \mathrm{im}(\partial_{i+1}),\\
    Z_i &= \mathrm{ker}(\partial_i).
\end{align*}
In our $2\times 2$ torus example,
\begin{align*}
B_1 &= \langle \partial \sigma_{10}, \partial \sigma_{01}, \partial \sigma_{11}\rangle\\
Z_1 &= \langle \gamma_0, \gamma_1, \partial \sigma_{10}, \partial \sigma_{01}, \partial \sigma_{11}\rangle,
\end{align*}
where $\gamma_0 = \ell_{000}+\ell_{010}$ and $\gamma_1 = \ell_{001}+\ell_{101}$.
The boundary group is literally generated by plaquettes boundaries, and the cycle group consists of chains without boundaries. 

Due to (\ref{eq:homology}) we always have $B_1\subset Z_1$,
geometrically showing that a boundary is always a cycle. However, a cycle is not necessarily a boundary, e.g. $\gamma_1, \gamma_2 \not\in B_1$. The difference is characterized by the homology group
\begin{displaymath}
    H_i = Z_i / B_i,
\end{displaymath}
In our $2\times 2$ torus example, 
$H_1 = \langle \gamma_1, \gamma_2\rangle$.

The rank of homology groups are Betti numbers,
$$
    b_i = \dim H_i.
$$
For \Z2LGT on the torus, $b_2 = 1$ is the number of dependent plaquette operators and $b_1 = 2$ is the number of topological qubits.

Cohomology emerges naturally when considering $\mathbb{Z}_2$ valued functions over chain groups, the cochain groups
$$C^i = \mathrm{Hom}(C_i, \mathbb{Z}_2).$$ 
The cochain groups are generated by characteristic functions over the chain groups, e.g.
$C^2 = \langle \mathbf{1}_\sigma\rangle_{\sigma\in\Sigma}$, $C^1 = \langle\mathbf{1}_{\ell}\rangle_{\ell\in\Lambda}$,
and $C^0 = \langle \mathbf{1}_{v}\rangle_{v\in V}$, where the characteristic functions $\mathbf{1}_{\sigma}$ are defined by  Kronecker pairing 
$$\langle \mathbf{1}_\sigma, \tau\rangle = \mathbf{1}_\sigma(\tau) = \delta_{\sigma\tau}.$$

Cochain groups are organized by coboundary operators $\delta^0, \delta^1$, or $\delta$ when the context is clear,
\begin{align*}
    \delta^0 \mathbf{1}_v &= \sum_{\ell: v\in\partial\ell} \mathbf{1}_\ell,\\
    \delta^1 \mathbf{1}_\ell &= \prod_{\sigma:  \ell\in\partial \sigma} \mathbf{1}_\sigma,
\end{align*}
giving rise to the cochain complex
$$
C^0 \stackrel{\delta^0}{\longrightarrow}
C^1 \stackrel{\delta^1}{\longrightarrow}
C^2,\quad \delta^1\circ\delta^0 = 0.
$$

The cohomology group is
$$
    H^i = Z^i / B^i = \mathrm{ker}(\delta^i) / \mathrm{im}(\delta^{i-1}).
$$
The cohomology group for our $2\times 2$ torus is $H^1 = \langle \gamma^1, \gamma^2\rangle$, 
where $\gamma^1 = \mathbf{1}_{\ell_{000}} + \mathbf{1}_{\ell_{100}}$ and $\gamma^2 = \mathbf{1}_{\ell_{001}} + \mathbf{1}_{\ell_{011}}$.

There is a natural isomorphism $C_i\cong C^i$ by identifying a geometric element $\sigma$ with its characteristic function $\mathbf{1}_\sigma$. This induces the adjoint operator $\partial^*$ that acts in the cochain group but behaves like the boundary operator, 
$$
\partial^* \mathbf{1}_\sigma = \sum_{\ell\in \partial\sigma} \mathbf{1}_\ell.
$$

The Laplacian operator is defined as
$$\Delta^i = \partial^*_{i+1} \delta^i + \delta_{i-1}\partial^*_i.$$
The null space of the Laplacian operator $\mathrm{ker}(\Delta^i)$ consists of $i$-dimensional harmonic fields allowed by the underlying topology. The combinatorial Hodge decomposition renders a deep understanding of these fields
$$
    C^i = \mathrm{im}(\delta^{i-1})\oplus \mathrm{ker}(\Delta^i)\oplus \ker{\partial^*_{i+1}}.
$$

\end{document}